\documentstyle[amssymbols,epsf]{article}
\title {Integrable extensions of the rational and trigonometric
 $A_N$ Calogero
Moser potentials }
\author{J. Avan \thanks{L.P.T.H.E. Universit\'e Paris VI
 (CNRS UA 280), Box 126, Tour 16,
$1^{er}$ \'etage, 4 place Jussieu, F-75252 PARIS CEDEX 05} }
\date{September 1993}
\def\x {\stackrel {\textstyle \otimes}{,}}
\begin{document}
\begin{titlepage}
\renewcommand{\thepage}{}
\maketitle
\begin{abstract}
We describe the $R$-matrix structure associated with
 integrable extensions, containing both one-body and two-body
 potentials, of the
$A_N$ Calogero-Moser $N$-body systems. We
construct non-linear, finite
dimensional Poisson algebras of observables. Their $N
 \rightarrow \infty$ limit realize the infinite Lie algebras
Sdiff$({\Bbb R} \times S_1 )$ in the trigonometric case and
 Sdiff$({\Bbb R }^2)$ in the
rational case. It is then isomorphic to the algebra of
 observables constructed  in the two-dimensional
collective string field theory.
\end{abstract}
\vfill

PAR LPTHE 93-23 \hfill
Work sponsored by CNRS, France.

\end{titlepage}
\renewcommand{\thepage}{\arabic{page}}

\section{Introduction}

Recent studies have lead to a number of results on the
 Calogero-Moser  $N$-body classical
models \cite{Ca,Mo}. In particular the $R$-matrix structure
 for the Lax operator of the $A_N$ Calogero
models was derived in the case of rational and trigonometric
 two-body potentials \cite {AT}.
This structure was interpreted and generalized to other
 algebras in \cite{ABT}. It follows
from the realization of these models by hamiltonian
 reduction \cite{AbM} of free or harmonic
motions on a suitable symmetric space, introduced in \cite{OP}.
 It provides us with the complete algebraic
framework for integrability of the two-body
 potential Calogero-Moser models.

We here extend \cite{ABT} to integrable
 spinless Calogero models containing an extra one-body term.
 We restrict ourselves to
the $A_N$ case.
 Such generalizations for the rational and trigonometric case
were introduced and
studied in \cite{Ad,Ino83,Woj84,Pol,Pe92}.

The systems under consideration have the following Hamiltonians:
 \begin{eqnarray}
H = \sum_{i=1}^{N} \frac{p_i^2}{2m} + \sum_{i \neq j}^{N} v_2 (q_i - q_j) &+&
 \sum_{i=1}^{N} v_2(q_i) \nonumber\\
{\rm with}\; v_2 = \frac{1}{{\rm (sinh \; or \; sine)}^2 (q_i - q_j)} &;& v_1 =
 (c_1 e^{2q_i} + c_2 e^{-2q_i} + c_0)^2
\nonumber\\
{\rm and}\;  v_2 = \frac{1}{ (q_i - q_j) ^2} &;& v_1 = (aq_i^2 + b q_i  +
c_0)^2
\label{1aa}
\end{eqnarray}

Note that they are not the most general Calogero systems of this type since
integrability can be
proved directly when $v_1$ is {\it any} linear combination of the monomials
 involved in (\ref{1aa}) \cite{Ino83,Woj84}. However their
associated algebraic structures can
be derived from basic principles without much computational difficulties,
and this will be our major interest here.

We first describe the particular scheme of
Hamiltonian reduction from which these models naturally arise.This enables
us to derive
 their associated $R$-matrix
structure. It is in fact a modified structure involving an extra term in
the Poisson brackets of the Lax operators. It is however shown to imply the
existence of a canonical $R$-matrix structure of the usual form \cite{STS,Skl}.

 Using this extended $R$-matrix structure we then construct closed Poisson
algebras of
 functions on the phase space, naturally
including the commuting Hamiltonians, providing us with a purely algebraic
framework for
the eventual
explicit resolution of the models.

In order to proceed, we recall
two fundamental results of the theory of integrable systems
\cite{STS,Skl,BV,Ma}:
\vskip 5mm

{\bf Proposition 1 :}

Given a Lax operator $L$, valued in a Lie algebra ${\cal G}$
 and depending on
canonically conjugate phase space variables $ p_i , q_j $; its
 adjoint invariant functions $ \{ {\rm Tr} L^n \} $
Poisson-commute if and only if there exists an $R$-matrix in
 ${\cal G} \otimes {\cal G}$, a priori
depending on the phase space variables, such that:

\begin{equation}
\{ L \x L \} = [R \, , \, L \otimes {\bf 1}] - [R^{\pi} \, , \,
 {\bf 1} \otimes L]
\label{3}
\end{equation}
where $ L \equiv \sum_{ t_{\alpha} } \, l^{\alpha}
(p_i , q_j ) t_{\alpha}$
 ; $\{ L \x  L \} \equiv \sum \sum \{ l^{\alpha} , l^{\beta} \}
 t_{\alpha} \otimes t_{\beta}$; $R^{\pi}$ denotes the action on $R$ of
the permutation operator $\Pi$
of the two algebras in ${\cal G} \otimes {\cal G}$; $ \{t_{\alpha} \}$ is a
 basis of ${\cal G}$.
\vskip 5mm

{\bf Proposition 2 :}

If a Lax operator $L$ has an $R$-matrix (\ref{3}), any
 conjugated Lax operator $L^u = u(p,q) L u^{-1}(p,q)$
has a conjugated $R$-matrix:

\begin{eqnarray}
R^u &=& u \otimes u \; R \; u^{-1} \otimes u^{-1} +
 {\bf 1} \otimes u \; \{ u \x L \} \; u^{-1}
\otimes u^{-1} \nonumber\\
&+& 1/2 \; \left[ \{u \x u \} \, u^{-1} \otimes u^{-1} , {\bf 1} \otimes L
\right]
\label{5}
\end{eqnarray}

\section{Calogero models from Hamiltonian reduction}

The pattern to obtain integrable
 trigonometric/hyperbolic Calogero-Moser models
runs as follows \cite{ABT,OP,Pol}.
Starting from a Lie group $G$ with a subgroup $K$ such that $G/K$ be a
symmetric space,
one takes as phase space the cotangent bundle of $G/K$ suitably parametrized
by $N \times N$ matrices $(X,Y)$ with the symplectic form
 $ \Omega  = \, {\rm Tr}~ dX \wedge dY$.
The free motion described by the Hamiltonian:
$H = {\rm Tr} XYXY$, is Liouville-integrable due to the existence
 of Poisson-commuting quantities $ H^{(n)}
\equiv {\rm Tr} (XY)^n$ including the Hamiltonian
itself \cite{Ar}.

 There is a natural left action of $G$, and particularly
 its subgroup $K$, on $G/K$, which
is symplectic and has a momentum map $ \mu = [X,Y] $ in
 our coordinates. Explicit Hamiltonian
reduction is achieved by first introducing a Cartan
decomposition of $X = uQu^{-1}\, , \, u \in H
\, , \,Q \in {\rm exp} \, \Lambda $ where $\Lambda $ is
 the Cartan algebra of the symmetric space $G/K$,
and similarly conjugating $Y$ as $\tilde{Y} = u^{-1} Y u$
\cite{OP}.

We choose from now on $G= Sl(N, {\Bbb C})$ and $K= SU(N)$.
Fixing the value of the momentum
$\mu$ to  $\mu_0 = v^T v - {\bf 1} \, , \, v = ( 1 ,... 1)$
 and dividing out by the invariance group $SU(N-1) \times U(1)$ of this fixed
momentum
achieves the Hamiltonian reduction to the Calogero-Moser trigonometric two-body
model
 [5], leaving as canonical
 variables the set defined by $Q = {\rm diag \; exp} \, q_i$
and $\tilde{Y}_{ii} \equiv p_i \, {\rm exp} -q_i$. The off-diagonal part
of the Lax operator $L = Q\tilde{Y}$ is
completely determined once the momentum $ \mu = [X,Y] \equiv
[Q,\tilde{Y}]$ has been fixed.

This specific choice of $\mu_0$ made in \cite{ABT,OP} is crucial to
preserve the integrability properties of the model, as was first shown
 in \cite{Pol}: it guarantees that the variables
 $u$ and the quantities $[Q, \tilde{Y}]$
are canonically conjugate in the neighborhood of the surface
 $u = {\bf1}, \mu = \mu_0$.
Precisely $\Omega$ reads:

\begin{equation}
\Omega = {\rm Tr} \{ dQ \wedge d \tilde{Y} + du . u^{-1}
 \wedge d [ Q,\tilde{Y}]
+ du .u^{-1} \wedge \left[ Q,[du .u^{-1} , \tilde{Y}]
 \right] \}
\label{1a}
\end{equation}

and when $\mu = \mu_0$ as defined above the last term vanishes
exactly on the surface $u={\bf 1} \, , \, \mu = \mu_0$.
 Hence  we have:
\vskip 5mm

{\bf Proposition 3}

Given a set of operators $L(X,Y)$ on the initial phase space, such
that $L(uXu^{-1}  , uYu^{-1}) = u L(X , Y) u^{-1}$,
the Poisson structure of the reduced operators $L(Q,\tilde{Y})$
is obtained by conjugation of the original Poisson structure and
{\it subsequent} elimination of the extra
variables $u$ and $[Q,\tilde{Y}]$.
\vskip 5mm

It follows that if the original $L$ operators have $R$-matrix structures,
the reduced Lax operators will have $R$-matrix structures obtained by
conjugation of the initial structure according to Proposition 2 and subsequent
elimination of $u$ and  $[Q,\tilde{Y}]$. Moreover one has:
\vskip 5mm

{\bf Proposition 4}

If a set of adjoint-invariant functions on the large phase space $F_n (X,Y)$
realizes a closed Poisson algebra, the set of
functions $F_n (Q,\tilde{Y})$ on the reduced phase space realizes an isomorphic
algebra.
\vskip 5mm

This comes from the fact that adjoint-invariance automatically
 eliminates $u$ as a
relevant variable in the set of $F_n$.

The  construction of integrable two-body plus one-body potentials
 now follows from Propositions 3 and 4. We
introduce a pair of Lax operators $ XY \pm V(X)  \equiv L^{\pm}$
 where $V$ is some (finite)
Laurent series. The candidate Hamiltonian is chosen to be
 $H = {\rm Tr} \, L^+ L^-$
As follows from Proposition 4 ,
 if the original Hamiltonian $H = {\rm Tr}(XYXY - V^2(X))$ is
 integrable, and the
 commuting action variables are invariant functions of $X$ and
$Y$, the reduction procedure
  immediately guarantees
that the reduced Calogero-type system will be integrable. We consider the set
$ \{ {\rm Tr} (L^+ L^-)^n \} $ as natural candidate action-variables, and
investigate under which condition they
will commute by deriving the Poisson structure of
 the operator
$ L^+ L^- $.

{\bf Remark:} When one considers  non simply-laced Lie
 algebras instead of $A_N$,
Calogero models generically contain one-body plus two-body
terms, related to the structure
of the roots of such algebras. These models have a Lax representation
with a single Lax
operator; the Hamiltonians have additional terms
$1/sinh^2 (q_i + q_j)$ \cite{OP,Ino} \footnote{These references
 were brought to our attention by
E. K. Sklyanin}.  It is possible that the more general models studied in
\cite{Ino83,Woj84} are related to such algebras.

\section{Extended trigonometric potentials}

We first compute the Poisson structure
 for $L^+ , L^- $ and $L^+ L^-$.
For $V(X) = X^n , n>0$ one has:

\begin{eqnarray}
\{ L^{\pm} \x L^{\pm} \} &=& [{\cal C}^{sl(N)}
,L^{\pm} \otimes {\bf 1}-{\bf 1}\otimes L^{\pm}] \nonumber \\
\{ L^{+} \x L^{-} \} &=& [{\cal C}^{sl(N)}
,L^{+} \otimes {\bf 1}-{\bf 1}\otimes L^{-}]\nonumber \\
&+& \sum_{m=1}^{n} {\cal C}^{sl(N)} (X^{n-m} \otimes X^m + X^{m}
 \otimes X^{n-m}) \label{7}
\end{eqnarray}
Here ${\cal C}^{sl(N)}$ is the quadratic Casimir operator of $sl(N, {\Bbb C})$
represented as :
$ {\cal C}^{sl(N)} = \sum_{i \neq j} e_{ij} \otimes e_{ji}
+ \sum_{i} e_{ii}\otimes e_{ii}$

 From (\ref{7}) and denoting ${\cal R} = {\bf 1} \otimes L^+ \,
 {\cal C}^{sl(N)} +{\cal C}^{sl(N)}
 L^- \otimes {\bf 1}$ one has:

\begin{eqnarray}
\{ L^{+} L^{-} \x  L^{+} L^{-} \} = [{\cal R} , L^{+} L^{-}
\otimes {\bf 1}] &-&
[{\cal R}^{\pi} ,{\bf 1} \otimes L^{+} L^{-}] \nonumber \\
 +{\cal C}^{sl(N)} \left\{ \sum_{m=1}^{n} L^+ (X^{n-m} L^- \otimes
 X^m \right. &+& L^+ X^{m} L^- \otimes X^{n-m})  \nonumber\\
-  (X^{n-m} \otimes L^+ X^m L^- &+& \left. X^{m} \otimes L^+  X^{n-m}
 L^- ) \right\}
\label{8}
\end{eqnarray}

For $n<0$, one gets similar formulas where the summation goes from
 $m=1$ to $-n$ and the
extra terms  in (\ref{7}, \ref{8}) become $-{\cal C}^{sl(N)}(X^{n+m-1}
 \otimes X^{-m+1} + X^{-m+1}
\otimes X^{n+m-1})$. Finally
if $V$ is a sum of monomials, every monomial contributes as one such
  sum multiplied by
 ${\cal C}^{sl(N)}$.

 At this point it is important to recall the following  property
 of ${\cal C}^{sl(N)}$
\begin{equation}
\forall a,b \in {\cal G} \equiv sl(N, {\Bbb C}) \; , \;
 {\rm Tr}_{{\cal G} \otimes {\cal G}} \;
  {\cal C}^{sl(N)} \: a \otimes b =
{\rm Tr}_{\cal G} \;  ab
\label{9}
\end{equation}
The following result is a consequence of (\ref{9}):
\vskip 5mm

{\bf Proposition 5 :}

If a Lax operator $L$ has a Poisson bracket structure
 of the extended form:

\begin{equation}
\{ L \x L \} = [R, L \otimes {\bf 1}] - [R^{\pi},
 {\bf 1} \otimes L]
+{\cal C}^{sl(N)} \sum_{i=1}^{k} (a_i \otimes b_i -
 b_i \otimes a_i)
\label{10}
\end{equation}

and if for all $i=1 \cdots k, a_i = L^p~{\rm or}~ b_i =
 L^p ~{\rm for~ some~ value~ of~ } p $ in ${\Bbb Z}$
then $\forall q,m  \in {\Bbb Z} ,~ \{ {\rm Tr} L^q , {\rm Tr}L^m \} = 0 $
and $L$ has therefore a genuine $R$-matrix structure
according to Proposition 1.
\vskip 5mm

Explicit derivation of this canonical $R$-matrix structure is a question which
we do not wish to address here. In fact the Poisson structure (\ref{10})
is perfectly suitable for the purpose of proving integrability (as shown in
Proposition 5)
and deriving the algebraic structure of observables as will soon be clear.
\vskip 5mm

If now terms of the form ${\cal C}^{sl(N)} (a_i \otimes
 b_i - b_i \otimes a_i)$ occur
with neither $a_i$ nor $b_i$ equal to a power of $L$, they
 will contribute to the Poisson bracket
$\{ {\rm Tr} L^p , {\rm Tr}L^q \}$ as Tr $ (L^{p-1}aL^{q-1}b -
L^{q-1}aL^{p-1}b) $ which does not
vanish a priori. Hence such a Lax operator does not in general
lead to an integrable hierarchy
of Hamiltonians, and may not have an $R$-matrix structure.

 From this result and examining (\ref{8}) we deduce:

a) For $n = 1,0,-1$ and linear combinations thereof
($L^\pm = XY \pm c_1 X \pm c_2 X^{-1} \pm c_0 {\bf 1}$),
 the Lax operator $L =  L^{+} L^{-}$ has an $R$-matrix
 structure. The Hamiltonian $H = {\rm Tr } L^{+} L^{-} $ is
therefore integrable; the action
variables are the higher traces $H^{(m)} = {\rm Tr } (L^{+} L^{-})^m $.
After Hamiltonian reduction, this gives rise to a set of
 two-body plus one-body hyperbolic
or trigonometric Hamiltonians of the form (redefining $q_i \leftarrow 2q_i$) :

\begin{eqnarray}
H \; = \; \sum_{i=1}^{N} \frac{p_i^2}{2m} &+& \sum_{i \neq j = 1}^{N}
 \frac{1}{({\rm sine~or~sinh})^2 (q_i - q_j)} \nonumber\\
+\sum_{j =1}^{N} c^2_1 \, e^{4(i)q_j} +  c^2_2 \, e^{-4(i)q_j} &+&
2c_1 c_0 e^{2(i)q_j} + 2 c_2c_0 e^{-2(i)q_j}
\label{11}
\end{eqnarray}

 This case was described in \cite{Pol}, where use was made
of the Poisson structure of the Lax matrix $L \pm c_1 X \pm c_2 X^{-1}$
providing a first example of our general structure (\ref{8}).

b) For all other exponents $n$, $L =  L^{+} L^{-}$ does not
 have a priori an $R$-matrix structure. This does not eliminate
definitely  corresponding extensions of the Calogero potentials;
 however the occurrence of terms
${\cal C}^{sl(N)} (a_i \otimes b_i - b_i \otimes a_i) $ will plague
 every attempt at defining higher conserved Hamiltonians
of the form ${\rm Tr~ Polynomials}~(L^{+} L^{-})$.

Finally the conjugation formula in Proposition 3 can be extended to a Poisson
bracket
 structure of the form (\ref{7}), where the supplementary
term $\sum {\cal C}^{sl(N)} \cdots $ is simply conjugated by $u \otimes u$.
 This leads to a Poisson bracket structure for the
Lax operators of  extended Calogero models of the form

\begin{eqnarray}
\{ L^{\pm} \x L^{\pm} \} &=& [{\cal R}
,L^{\pm} \otimes {\bf 1} ]-[ {\cal R}^{\pi} , {\bf 1}\otimes L^{\pm}] \nonumber
\\
\{ L^{+} \x L^{-} \} &=& [{\cal R}
,L^{+} \otimes {\bf 1}]-[{\cal R}^{\pi},{\bf 1}\otimes L^{-}]\nonumber \\
&+& \sum_{m=1}^{n} {\cal C}^{sl(N)} (Q^{n-m} \otimes Q^m + Q^{m}
\otimes Q^{n-m}) \label{12}
\end{eqnarray}

for any $V(X) = X^n, n \in {\Bbb N}$. The signs and indices are changed
accordingly for
 $V(X) = X^{-n}$.  The $R$-matrix
in (\ref{12}) is the same as for the pure two-body case \footnote{The
discrepancy with formulas
in \cite{ABT,AT} is due to an extra conjugation of $L$ by Diag exp $q_i /2 $ }:

\begin{eqnarray}
R &=& \sum_{i \neq j} {\rm coth} (q_i - q_j) e_{ij}
 \otimes e_{ji} \nonumber\\
 &+& 1/2 \sum_{i \neq j} ( {\rm coth} (q_i - q_j) + 1) (e_{ii} + e_{jj})
\otimes e_{ij}
  + \sum_{i} e_{ii} \otimes e_{ii} \label{1b}
\end{eqnarray}

\section{Algebra of observables}

We shall first examine the  case
 where $V$ is a {\it monomial} $V(X) = X$ or
$V(X) = X^{-1}$.
Since $ [{\tilde Y} , Q] = \mu_0$, one has

\begin{eqnarray}
[L^{+} , L^{-}] &=& \mu_0 (L^+ - L^-)  \, ; \, V = Q  \nonumber\\
 \noindent
[ L^{+} , L^{-} ] &=& -(L^+ - L^-)\mu_0 \, ; \, V = Q^{-1}
\label{13}
\end{eqnarray}

This allows us to introduce a notion of normal-ordered
 observables.
\vskip 5mm

{\bf Proposition 6:}

 For any monomial trace Tr$(L_1 \cdots L_n)$ where $L_i = L^\pm$
and $V(Q) = Q$ or $Q^{-1}$, there exists a polynomial
expression of this trace in terms of normal ordered traces
 Tr$(L^+ )^n (L^-)^m$.
\vskip 5mm

The proof runs as follows.
Denoting the length of the monomials by $l(A)$ one first checks
the statement for $l(A) = 1,2$ in which case it is trivial.
One then assume that the normal ordering is proved for $(L_1 \cdots L_n)$
 until $n=n_0 + 1$. Consider now Tr$(L_1 \cdots L_{n_0 +2})$.
Normal-ordering is achieved by commuting every
$L^+$ with every $L^-$ on its left, leading to

\begin{equation}
{\rm Tr}(L_1 \cdots L_{n_0 +2}) = {\rm Tr}(L^+ )^a (L^-)^b \, + \,
 \sum {\rm Tr}([L^+ , L^- ]L'_1 \cdots L'_{n_0})
\label{14}
\end{equation}

 One can rewrite every supplementary term in (\ref{14}) as:

\begin{eqnarray}
{\rm Tr}([L^+ , L^- ]L'_1 \cdots L'_{n_0}) &=& {\rm Tr}[L^+ ,
L^- L'_1 \cdots L'_{n_0}] \nonumber\\
 &+& \, \sum {\rm Tr}([L^+ , L^- ]L''_1 \cdots L''_{n_0})
\label{15}
\end{eqnarray}

where $L''_1 \cdots L''_{n_0}$ denotes a reordering of $L'_1 \cdots
 L'_{n_0}$ each time $L^+$ commutes with one $L^-$ in the
expansion of Tr $[L^+ , L^- L'_1 \cdots L'_{n_0}]$ . Hence  $L''_1 \cdots
 L''_{n_0}$ can be reexpressed as
 $L'_1 \cdots L'_{n_0}$ up to extra commutators, leading to the equation:

\begin{eqnarray}
0 = {\rm Tr}[L^+ , L^- L'_1 \cdots L'_{n_0}] &=& (1+k){\rm Tr}([L^+ ,
 L^- ]L'_1 \cdots L'_{n_0})\nonumber\\
&+& \sum {\rm Tr} ( [L^+ , L^-] A [L^+ , L^-] B)
\label{16}
\end{eqnarray}

In (\ref{16}) $k$ denotes the number of $L^-$ terms in $L'_1 \cdots
 L'_{n_0}$; $A$ and $B$ are monomials in $L^\pm$
of total length $l(A) + l(B) = n_0 - 2$. Using (\ref{13})
 all remaining terms in (\ref{16})
have the structure (respectively for the potential  $X^{-1}$ or $X$):
${\rm Tr} (\mu_0 (L^+ - L^-) A \mu_0 (L^+ - L^-) B)$ or
 $((L^+ - L^-)\mu_0 A (L^+ - L^-)\mu_0 B)$

Since $\mu_0 = \sum_{i \neq j} e_{ij}$ one has the remarkable property:

\begin{equation}
{\rm Tr} ((\mu_0 + {\bf 1}) A (\mu_0 + {\bf 1}) B) =  {\rm Tr}
((\mu_0 + {\bf 1}) A) {\rm Tr} ((\mu_0 + {\bf 1}) B)
\label{18}
\end{equation}

hence one gets:

\begin{eqnarray}
&{\rm Tr}& (\mu_0 (L^+ - L^-) A \mu_0 (L^+ - L^-) B) = \nonumber\\
  &{\rm Tr}& ((\mu_0 + {\bf 1})(L^+ - L^-) A)
 {\rm Tr} ((\mu_0 + {\bf 1})(L^+ - L^-) B)
\nonumber\\
- &{\rm Tr}& ( \mu_0 (L^+ - L^-) B(L^+ - L^-) A) - {\rm Tr}
 ( \mu_0 (L^+ - L^-) A(L^+ - L^-) B) \nonumber\\
-  &{\rm Tr}& (L^+ - L^-) B(L^+ - L^-) A)
\label{19}
\end{eqnarray}

Since $l(A) + l(B) = n_0 - 2$, every monomial trace on the
 right hand side of (\ref{19}) falls under the recursion
hypothesis. In particular one replaces the expression $\mu_0 (L^+ - L^-)$
by the commutator $[L^+ , L^-]$ thereby generating in (\ref{19})  monomials of
length
at most $n_0 +1$ to which the recursion hypothesis applies.

This demonstration is here specifically given for
 $V(X) =X^{-1}$. In the other case
one  considers expressions of the form Tr $(L^+ - L^-)\mu_0
 A (L^+ - L^-)\mu_0 B)$; one simply needs to apply cyclicity of the
trace operation in order to get terms of the form (\ref{19}). The
conclusion is preserved.

This ends the proof of the recursion at the level $n_0 +2$. Hence
the normal-ordering Proposition is proved.

We now derive the algebra of observables
 $\{ W_n ^m = {\rm Tr} (L^+)^n (L^-)^m \}$.
 From the Poisson structure (\ref{12}) it follows that (for $V(X) = X$):

\begin{eqnarray}
\{ W_n ^m , W_p ^q \} &=& - \sum_{a,b} {\rm Tr} (L^+ - L^-) \{ (L^-)^{a-1}
(L^+)^{n}(L^-)^{m-a}(L^+)^{b-1}(L^-)^{q}(L^+)^{p-b}  \nonumber\\
 &+& (L^+)^{b-1}
(L^-)^{q}(L^+)^{p-b}(L^-)^{a-1}(L^+)^{n}(L^-)^{m-a} \} \nonumber\\
&+& \sum_{c,d} {\rm Tr} (L^+ - L^-) \{ (L^+)^{c-1}
(L^-)^{m}(L^+)^{n-c}(L^-)^{d-1}(L^+)^{p}(L^-)^{q-d} \nonumber\\
 &+& (L^-)^{d-1}
(L^+)^{p}(L^-)^{q-d}(L^+)^{c-1}(L^-)^{m}(L^+)^{n-c} \}
\label{19a}
\end{eqnarray}

For $V(X) = X^{-1}$ the global sign changes. The normal-ordering
 mechanism now leads to :

\begin{eqnarray}
\{ W_n ^m , W_p ^q \} \, &=& \, (mp-nq) \{W_{n+p}^{m+q-1} +
 W_{n+p-1}^{m+q} \} \nonumber\\
&+& {\cal P} (W^a_b \,,\, a\leq n+p-2\, ;\, b \leq m+q-2 )
\label{20}
\end{eqnarray}

\noindent where ${\cal P}$ can be explicitely computed using the previous
recursion procedure.

The Poisson algebra of observables therefore realizes a structure
 evoking a non-linear deformation of a $W_\infty $
algebra on a cylinder. Indeed the linear terms in (\ref{20}) are
 reproduced as the Poisson bracket structure of the
set:
$  W^n_m \equiv (p+ e^{iq})^n (p-e^{iq})^m $ with $ \{ p,q \} = 1$.

One must emphasize  however that the normal-ordering procedure is
 {\it not univocally} defined:
 another ordering of the steps in the recursion may lead to a different
 but equivalent expression
in term of the explicit matrices given their non-generic features.
 Moreover, there exists necessarily an ideal
of strictly zero polynomials of the adjoint-invariant quantities
 Tr$(L^+)^n (L^-)^m $, since when the matrix
$L$ is of size $N$ the quantities Tr$(L^p)\, , \, p>N$ are polynomial
 functions of the independent quantities
Tr$(L^p)\, , \, p \leq N$.  Hence (\ref{20}) is in fact an
ambiguously defined expression as long as the quotient
by this trivial ideal is not explicitely achieved. In particular
 this algebra can only have a finite number
of generators as long as $N$ is finite.

The infinite-dimensional
linearized limit can however be well-defined. First of all,
the recursive proof shows that the next-to-longest term
in the normal-ordering procedure of Tr$(L_1 \cdots L_m)$ has a total
 length $m-1$.  Hence if one renormalizes:
$\overline{W^p_q} \equiv N^{-p-q+1} W^p_q $
one gets:

\begin{equation}
\{ \overline{W_n ^m} ,\overline{ W_p ^q} \} \, = \, (mp-nq)
 \{\overline{W_{n+p}^{m+q-1}} +
 \overline{W_{n+p-1}^{m+q}} \} + N^{-1} \: {\cal P} (\overline{W^a_b})
\label{22}
\end{equation}

Moreover we expect that when $N$ becomes large, at least all
 normal-ordered monomials $\overline{W_n ^m}$ up to
a value of $n+m$ of order $N$ will be algebraically independent.
The $N \rightarrow \infty$
limit of (\ref{22}) is then well-defined and
 coincides with Sdiff $(S_1 \times {\Bbb R})$.

In particular, recalling that $H \, = \, {\rm Tr} \, L^+ L^- = W^1_1$  one has
:

\begin{equation}
 \{ H\, ,\, W^n_m \} \, = \, (n-m) \{ W^m_{n-1} \, + \, W^{m-1}_n \}
+ 0( N^{-2}) ...
\label{23}
\end{equation}

This does not allow to solve the eigenfunction equation
 $\{ H,{\cal O} \} = \epsilon {\cal O}$ by setting $\cal{O}$
to be a finite linear combination of normal-ordered observables.
 A related discussion
appears in \cite{AJ} for collective string
 field theory.
\vskip 5mm

Let us consider now the case when both monomials $X$
and $X^{-1}$ are retained
in the potential (keeping the term $c_0 {\bf 1}$ does not modify our
generating set of observables).
The Poisson brackets (\ref{7}) cannot be rewritten purely in terms
of $L^+ - L^-$ due to the change of relative sign between $X$
 and $X^{-1}$
contributions. To get a closed Poisson algebra, one needs a priori to introduce
a three-index set of observables generated by terms of the
form
Tr $(Q^n (L^+)^m(L^-)^p)$.

First of all the problem of normal ordering must be revisited.
The relevant commutators are:

\begin{equation}
[L^+ \, , \, L^-] =  c_2 \mu_0 Q^{-1} - c_1 Q \mu_0 \, ; \,
[Q \, , \, L^\pm] = Q \mu_0 \, ; \, [Q^{-1} \, , \, L^\pm] =  -\mu_0 Q^{-1}
\label{1c}
\end{equation}

The normal-ordering Proposition states:
\vskip 5mm

{\bf Proposition 7}

Any monomial trace $ Tr( A_1 \cdots A_n )$ with $A_i \in \{ L^+ , L^- , Q ,
Q^{-1} \}$
can be reexpressed as a polynomial of normal-ordered traces $Tr
Q^m(L^+)^n(L^-)^p$.
\vskip 5mm

The proof involves a double recursion procedure. One denotes by $n(L^{\pm})$
the number of
such generators in a given monomial.

\indent {\bf Step 1}: Proof of Proposition 7 for $Tr( A_1 \cdots A_N )$, $n(L^-
) = 0$ .

 For $N=1$ normal-ordering is obvious.
 Assume it is proved
up to order $N_0$ (recursion hypothesis R1). Take $Tr( A_1 \cdots A_{N_0 +1}
)$.
 Normal-ordering this expression involves
commutators of the form (\ref{1c}), hence one has:

\begin{eqnarray}
Tr ( A_1 \cdots A_{N_0 +1} ) = Tr Q^p (L^+)^{n(L^+)} &+& \sum Tr Q \mu_0 A_1
 \cdots A_{N_0 -1}
\nonumber\\
&+& \sum Tr  \mu_0 Q^{-1} A_1 \cdots A_{N_0 -1}
\label{25a}
\end{eqnarray}

Any term of the form $Tr Q \mu_0 A_1 \cdots A_{N_0 -1}$ can be rewritten as
in (\ref{15}):

\begin{eqnarray}
0 = Tr [Q, L^+  A_1 \cdots A_{N_0 -1}] &=& (1 + n(L^+)) Tr Q \mu_0 A_1 \cdots
A_{N_0 -1}
\nonumber\\
&+& \sum
 ({\rm reordering  \; corrections } )
\label{25b}
\end{eqnarray}

The corrections take the form $ Tr Q \mu_0 A Q \mu_0 B$ and
 $ Tr Q \mu_0 A  \mu_0 Q^{-1} B$ with a length $l(A) + l(B)$ at most $N_0 -3$.
Using the decoupling property of $\mu_0$ (\ref{18}) generates:

1) from $ Tr Q \mu_0 A Q \mu_0 B$ : $ Tr Q  A Q  B$ ; $ Tr Q \mu_0 A Q  B$;
$ Tr Q \mu_0 B Q  A$; $ Tr Q \mu_0 A$; $ Tr  Q \mu_0 B$. Reexpressing
$Q \mu_0$ as $[Q, L^+]$ turns all such terms into monomials of length
$N_0$ or less to which R1 applies.

2) from $ Tr Q \mu_0 A  \mu_0 Q^{-1} B$ : $ Tr Q  A   Q^{-1} B$ ;
 $ Tr Q \mu_0 A  Q^{-1} B$;  $ Tr Q  A  \mu_0 Q^{-1} B$;  $ Tr \mu_0 A $;
 $ Tr  \mu_0 Q^{-1} B Q$. Reexpressing $Q \mu_0$ and $\mu_0 Q^{-1}$ as
commutators,
and $\mu_0$ as $ Q^{-1}[Q, L^+]$ turns all these terms into monomials of length
$N_0$ or less to which R1 applies.

 Hence (\ref{25b}) leads to an explicit normal-ordered
expression for any term $Tr Q \mu_0 A_1 \cdots A_{N_0 -1}$. A similar argument
holds for
$Tr  \mu_0 Q^{-1} A_1 \cdots A_{N_0 -1}$. Therefore (\ref{25a}) gives a
normal-ordered expression
for any monomial of length $N_0 +1$. This ends the proof of Proposition 7 for
$n(L^- ) = 0$.

\indent {\bf Step 2} : Assume that Proposition 7 holds for any monomial with
$n(L^-)$ up
 to $n_0$ (recursion hypothesis R2).

Consider now $n(L^-) = n_0 +1$, and a general monomial
in this subset $Tr A_1 \cdots A_N , N \geq  n_0 +1$. For $N = n_0 +1$
normal-ordering is
already achieved! Assume now that it has been proved for $n(L^-) = n_0 +1$ and
up to a total
length $N_0$ (recursion hypothesis R1). Take a monomial $Tr A_1 \cdots A_{N_0
+1}$.
 The general normal-ordering equation
(\ref{25a}) still holds since all commutators are of the form (\ref{1c}). The
residual terms
$A_1 \cdots A_{N_0 -1}$ contain at most $n_0 +1$ terms $L^-$.

Rewrite now an equality of the type (\ref{25b}). Since $Q$ has identical (up to
a sign)
non-vanishing commutators with both $L^+$ and $L^-$ one has now a more
complicated expression:

\begin{eqnarray}
0 = Tr  [Q, L^+  A_1 \cdots A_{N_0 -1}] &=& (1 + n(L^+)) Tr Q \mu_0 A_1 \cdots
A_{N_0 -1}
\nonumber\\
+ \sum Tr Q \mu_0 A'_1 \cdots A'_{N_0 -1} &+& {\rm
  reordering \;  terms }
\label{25c}
\end{eqnarray}

The reordering terms (coming from $Tr Q \mu_0 A_1 \cdots A_{N_0 -1}$ )
 contain at most $n_0 + 1$ $L^-$ terms.
 Moreover they have exactly the same
form as in Step 1. They can therefore  be
rewritten as normal-ordered polynomials due to hypothesis R1 and R2.

Now extra contributions $Tr Q \mu_0 A'_1 \cdots A'_{N_0 -1}$ have appeared in
(\ref{25c})
where $A'_1 \cdots A'_{N_0 -1}$ contains $1 + n(L^+)$ terms $L^+$ and {\it at
most}
$n_0$ terms $L^-$. Replacing in these terms the original $[Q, L^-] \equiv Q
\mu_0$
by $[Q, L^+]$ which is identical, turns them into monomials with at most
$n(L^-) = n_0$.
The recursion hypothesis R2 thus holds for these monomials. Hence all terms
in the second line of (\ref{25c}) can be normal-ordered, and so
can $Tr Q \mu_0 A_1 \cdots A_{N_0 -1}$. Consequently the normal-ordering
proposition is
proved for $n(L^-) = n_0 +1 , N = N_0 + 1$. Hence it is proved by recursion for
 $n(L^-) = n_0 +1$,
and any total length N. Together with Step 1, this finally proves
 it by recursion for any $n(L^-)$.
\vskip 5mm

Two degeneracy relations must now be implemented in order to eliminate
redundant generators:
\begin{eqnarray}
 L^+ - L^- = 2 c_1 Q - 2 c_2 Q^{-1} &\Rightarrow & Q^{-1} = 1/2c_2 (L^+ - L^- -
2 c_1 Q)
\nonumber\\
Q Q^{-1} \! = {\bf 1} = \frac{Q}{2c_2} (L^+ \! - \! L^- \! - \! 2 c_1 Q)
 \!\!\! \! & \Rightarrow& \!\!\!\!  Q^2 =
-\frac{c_1}{c_2}(1- \frac{1}{2c_2} (QL^+ \! - \! QL^-) \label{25d}
\end{eqnarray}

 It follows
that one shall consider as algebra of normal-ordered generators
the set $ \{ Tr Q^n (L^+)^m (L^-)^p, n \in \{ 0,1 \} , m,p \in {\Bbb Z} \} $.

 In order to get the Poisson structure of this algebra, it is here
 easier to use Proposition 4: one computes the Poisson brackets of observables
on the full phase space
$ \{ Tr X^n (L^+)^m (L^-)^p, n \in \{ 0,1 \} , m,p \in {\Bbb Z} \} $.
 From Proposition 4 the Poisson algebra of reduced observables
 $ \{ Tr Q^n (L^+)^m (L^-)^p, n \in \{ 0,1 \} , m,p \in {\Bbb Z} \} $ is
isomorphic to
it. One  then applies the normal-ordering procedure.

Although the finite $N$ algebra is quite complicated and again ambiguously
defined, we can easily derive its leading linear order -i.e. the
large $N$ limit.
It is obtained by reordering all terms in the full phase space algebra as
Tr $(X^n (L^+)^m(L^-)^p)$ and
dropping all induced commutators.
Hence, using (\ref{7}) and
 $\{ X \x L^\pm \} = {\bf 1} \, \otimes X \,
{\cal C}^{ sl(n)}$
one ends up with a general formula for $n_1,n_2 \in \{ 0,1 \}$:

\begin{eqnarray}
&&\left\{ \right. \! {\rm Tr} (Q^{n_1} (L^+)^{m_1}(L^-)^{p_1})
 \, , \, {\rm Tr} (Q^{n_2} (L^+)^{m_2}(L^-)^{p_2})
\left. \right\} \nonumber\\
&=& ((n_1 - p_1)m_2 - (n_2 - p_2)m_1) {\rm Tr} (Q^{n_1 + n_2}
 (L^+)^{m_1 + m_2 -1}(L^-)^{p_1 + p_2}) \nonumber\\
&+& ((n_1 - m_1)p_2 - (n_2 - m_2)p_1) {\rm Tr} (Q^{n_1 + n_2}
 (L^+)^{m_1 + m_2 }(L^-)^{p_1 + p_2 - 1}) \nonumber\\
&+& 2c_1 (m_1p_2 - m_2p_1) {\rm Tr} (Q^{n_1 + n_2 +1}
(L^+)^{m_1 + m_2 -1}(L^-)^{p_1 + p_2 -1})
\nonumber\\
&+& \cdots {\rm lower ~ orders}
\label{1e}
\end{eqnarray}
One must also implement the second degeneracy condition from (\ref{25c}) and
eliminate
$Q^2$ wherever it appears in (\ref{1e}) in order to get the algebraic structure
in term
of the independent generators.

 The large $N$ limit is then achieved by redefining
 generators Tr $(Q^n (L^+)^m(L^-)^p) \rightarrow N^{n+m+p-1} $ Tr $ (Q^n
(L^+)^m(L^-)^p)$.
 (\ref{1e}) then leads to a lengthy linear algebraic structure  mixing both
types of
generators $Tr  (L^+)^n(L^-)^m$ and $Tr  Q(L^+)^n(L^-)^m$. We shall not write
it explicitely,
since anyhow
we lack for the moment an interpretation of this algebra
 in term of diffeomorphisms of a surface.
It is relevant for the study of the collective field
 theory corresponding to particular
unitary matrix models \cite{JS,Do}. Finally note that adding the term $ \pm c_0
{\bf 1}$
only modifies the degeneracy relations (\ref{25d}) by adding lower-order terms
which will
disappear from (\ref{1e}) after taking the large $N$ limit.

\section{Extended rational potentials}

Rational pure two-body potentials are obtained by normalizing
 the $q$ coordinates in $Q$ as $Q = {\rm exp}\, aq_i$
and sending $a$ to zero. Their original Lax operator is then $Y$.
 Extended potentials  arise from considering shifted Lax
 operators $L^\pm  \, \equiv \, Y \, \pm \, V(X)$.
The Poisson brackets for $L^\pm$ are essentially identical to
(\ref{7}) up to one power of $X$ in the extra
$\sum$. It follows from similar considerations that integrable
 potentials -for which the Poisson brackets
take the form (\ref{10})- are  $V(X) \, = \, X^2 \, ,\,
 X~{\rm and}~ {\bf 1}$. We shall restrict ourselves to the monomial cases.

The case $V(X) = X^2$ leads to an integrable quartic Hamiltonian:

\begin{equation}
H_0 = \sum_{i=1}^{n} \frac{p_i^2}{2m} \; + \; \sum_{i\neq j=1}^{n}
 \frac{1}{(q_i -q_j)^2}
\; + \; \sum_{i=1}^{n} g q_i^4.
\label{24}
\end{equation}

A similar problem arises in this case to define a
 normal-ordered set of observables.
Poisson brackets of $L^+$ with $L^-$ contain terms
${\cal C}^{sl(N)} \,( Q \otimes {\bf 1}
 + {\bf 1} \otimes Q)$ which cannot be rewritten as
 linear functions of
$L^+$ and $L^-$. This
here requires adding extra observables of the form
 Tr $QL_1 \cdots L_m$ in order to have
a closed Poisson algebra.
 Contrary to the trigonometric case this algebra of
observables will exhibit a structure of symmetric
algebra as $\{ {\rm Tr} QL_1 \cdots L_m \}
 \oplus \{ {\rm Tr} L_1 \cdots L_m \}$.

The relevant commutators for the normal ordering
procedure are:

\begin{equation}
[L^+ \, , \, L^-] = \mu_0 Q \, + \, Q \mu_0 \; ; \;
[Q \, , \, L^\pm ] = \mu_0 \label{toto3}
\end{equation}

A normal-ordering procedure expresses monomials of the form
$Tr A_1 \cdots A_N$ with $ A_i \in \{ L^{\pm}, Q\}$ as polynomials of $Tr Q^m
(L^+)^n (L^-)^p$
The commutatorstructure defined in (\ref{toto3}) again triggers a double
recursion proof:

\indent {\bf Step 1.} Normal-ordering is proved for $n(L^-) = 0$ on the same
lines as in the previous cases,
since any commutator decreases the length of the monomial by two units:

\begin{equation}
Tr A_1 \cdots A_N = Tr Q^m (L^+)^n + \sum Tr \mu_0 A_1 \cdots A_{N-2}
\label{29a}
\end{equation}
Then since $[Q \, , \, L^\pm ] = \mu_0$, one rewrites

\begin{equation}
0 = Tr [Q, L^+ A_1 \cdots A_{N-2}]= (1+n(L^+)) Tr \mu_0  A_1 \cdots A_{N-2} +
{\rm reordering \;  terms }
\label{29b}
\end{equation}
Reordering corrections have the form $Tr \mu_0 A \mu_0 B$ with $l(A) + l(B) =
N_0 -4$. Hence
the decoupling mechanism (\ref{18}) generates, after replacing $\mu_0$ by $[Q,
L^+]$, monomials
of maximal length $N_0 -2$ allowing the recursion to hold.

\indent {\bf Step 2.} Recursion proof on $n(L^-)$.

The normal-ordering hypothesis is assumed to be proved up to $n(L^-) = n_0$
(hypothesis R2).

Here one needs to be more careful when considering the reordering terms and the
replacement
of commutators in them since, contrary to all previous cases, the commutators
of fundamental generators here assume
qualitatively different forms $\mu_0$ or $\mu_0 Q \, + \, Q \mu_0$. This is
dealt
with by introducing a different recursion hypothesis :

For $n(L^-) = n_0 + 1$, we assume that the normal-ordering proposition is
proved BOTH
for terms of the form $Tr A_1 \cdots A_N$ and $Tr \mu_0 A_1 \cdots A_N$ for $N$
up to
$N_0$ (hypothesis R1). For $N= n(L^-)$ the proposition is trivial since $Tr
\mu_0 (L^{\pm})^m = 0$.

Take now $Tr A_1 \cdots A_{N_0+1}$. Normal-ordering of this expression
generates corrective
terms of the form $Tr \mu_0 A_1 \cdots A_{N_0}$ or even $Tr \mu_0 A_1 \cdots
A_{N_0-1}$ to
which the recursion hypothesis applies.

Now take $Tr \mu_0 A_1 \cdots A_{N_0 +1}$. It is rewritten as:

\begin{eqnarray}
0 = Tr [Q, L^+ A_1 \cdots A_{N_0 +1}] &=&  (1 + n(L^+)) Tr \mu_0 A_1 \cdots
A_{N_0 +1}
\nonumber\\
+ \sum Tr \mu_0 A'_1 \cdots A'_{N_0 +1} \!\!\! \! &+& \!\!\!\! \sum Tr
\mu_0 A \mu_0 B \; {\rm( reordering \;
corrections)}
\label{29c}
\end{eqnarray}

The term $Tr \mu_0 A'_1 \cdots A'_{N_0 +1}$, contributed to by commutators
$[Q, L^-]$,
contains at most  $n(L^-) = n_0 $; hence the recursion hypothesis R2 can be
applied
after reexpressing
$\mu_0$ as $[Q, L^+]$.

The corrective terms $Tr \mu_0 A \mu_0 B$ have a total length of monomials of
at most $N_0$;
They generate monomials $Tr AB$, $Tr A \mu_0 B$, $Tr \mu_0 A B$, $Tr \mu_0 A$,
$Tr \mu_0 B$
to which R1 applies globally.
Hence normal-ordering is proved for $n(L^-) = n_0 + 1$, $N = N_0 +1$; therefore
it is
proved by recursion for any $N$ and $n(L^-) = n_0 + 1$. Finally Step 1 and Step
2 guarantee
that the normal ordering procedure holds for all values of $n(L^-)$.
\vskip 5mm

One degeneracy relation exists here, namely
$ Q^2 = L^+ - L^-  $.  Hence one shall define the relevant
set of observables as $\{ V_m^n \equiv {\rm Tr} Q (L^+)^n(L^-)^m \, ; \,
W^n_m \equiv {\rm Tr} (L^+)^n(L^-)^m \}$.

We compute again the leading
order of the Poisson algebra by simply considering the reordered Poisson
algebra without the extra commutators
on the full phase space of $X,Y$. Recalling the basic Poisson bracket
$\{ X \x Y \} = {\cal C}^{sl(N)}$
we get a symmetric Lie algebra structure:

\begin{eqnarray}
\{ V_{m_1}^{n_1} , V_{m_2}^{n_2} \} &=& (2(n_1 m_2 - n_2 m_1 )
 +m_2 -m_1) V_{m_1 +m_2 -1}^{n_1 +n_2}
\nonumber\\
&-& (2(n_1 m_2 - n_2 m_1 ) +n_2 -n_1 ) V_{m_1 +m_2 }^{n_1 +n_2
-1 } + {\rm lower \; orders} \nonumber\\
 \noindent
\{ V_{m_1}^{n_1} , W_{m_2}^{n_2} \} &=& (2(n_1 m_2 - n_2 m_1 )
 +m_2) W_{m_1 +m_2 -1}^{n_1 +n_2}
\nonumber\\
&-& (2(n_1 m_2 - n_2 m_1 ) +n_2  ) W_{m_1 +m_2 }^{n_1 +n_2 -1}
 + {\rm lower \; orders} \nonumber\\
\noindent
\{ W_{m_1}^{n_1} , W_{m_2}^{n_2} \} &=& 4(n_1 m_2 - n_2 m_1 )
 V_{m_1 +m_2 -1}^{n_1 +n_2 -1}+ {\rm lower \; orders}
\label{toto2}
\end{eqnarray}

The large $N$ limit is achieved by redefining $V_m^n \rightarrow
N^{n+m-1}V_m^n$ and
 $W^n_m \rightarrow N^{n+m -3/2} W^n_m $.

The algebra of $V$ generators recalls quite closely the
algebra Sdiff$({\Bbb R} \times S^1)$ for
trigonometric potentials (\ref{22}), up to a shift of the $n$
indices by $1/2$.
\vskip 5mm

The case $V(X) = X$ leads to the original so-called Type-V potential
 \cite{OP}. In this case
 $[L^+ \, , \, L^- ] = \mu_0 $ . Here the normal-ordering procedure
 can be derived in an
 exactly similar way to the first case of trigonometric potentials.

The observables
are defined as $W^n_m \equiv \{ {\rm Tr} (L^+)^n (L^-)^m \} $.
 The algebra then reads:

\begin{equation}
\{ W_n ^m , W_p ^q \} \, = \, (mp-nq) W_{n+p-1}^{m+q-1} + {\rm lower
 ~orders}
\label{25}
\end{equation}

The large $N$ limit is there achieved
by defining $\overline{W^p_q} \equiv N^{(-p-q+2)} W^p_q $.

A major difference with the trigonometric case is that the leading terms of the
Poisson
 algebra, which survive in the large $N$ limit
to give the linearized algebra, realize
 the $W_\infty$ algebra Sdiff$({\Bbb R}^2)$. The $N \rightarrow \infty$ limit
of the algebra is therefore isomorphic to the algebra of observables
 for the collective string field theory considered
 in \cite{AJ}. This statement is the correct formulation of the
 identification indicated in  \cite{DJT} between
observables of the matrix model and observables of the collective theory.

Finally the linearized Poisson algebra (\ref{25}) is in fact
exact when $n=m=0,1$. This implies:
\vskip 5mm

{\bf Proposition 8:}

$W^n_m$ are eigenfunctions of the Calogero Hamiltonian flow with
 energy $n-m$.
\vskip 5mm

 The proof runs as follows. The Poisson bracket Tr $ \{ L^+ L^-
 \x (L^+)^n (L^-)^m \}$ only gets contributions
from the extra terms in (\ref{7}). Hence:

\begin{eqnarray}
&{\rm Tr}& \{ L^+ L^- \x (L^+)^n (L^-)^m \} = \sum_{a,b} {\rm Tr}
 \{ -L^+ \otimes (L^+)^{n-a}\,
 {\cal C}^{  sl(n)}
{\bf 1} \otimes (L^+)^{a-1}(L^-)^{m} \nonumber\\
&+& {\bf 1} \otimes (L^+)^{n}(L^-)^{m-b} {\cal C}^{ sl(n)} L^-
\otimes (L^-)^{b-1} \} \nonumber\\
&=&\sum_{a,b} {\rm Tr} \{ -(L^+)^a (L^-)^m (L^+)^{n-a} \, + \,
 -(L^-)^b (L^+)^n (L^-)^{m-b} \} \nonumber\\
&=& (m-n) {\rm Tr} (L^+)^n(L^-)^m
\label{26}
\end{eqnarray}

This construction was first used in \cite{OP2} although the proof
 is rather involved. The Type V potential
realizes a consistent discretization of the hermitian one-matrix
 model \cite{AJ2}.
The corresponding derivation of eigenoperators for the associated
 collective field Hamiltonian
was given in \cite{AJ}.
\vskip 5mm

{\bf Acknowledgements}

Part of this work was done at Brown University Physics Department under
 the CNRS-NSF Exchange Programme
CNRS AI 0693. I wish to thank Antal Jevicki for fruitful suggestions, and A.P.
Polychronakos for bringing
to my attention references \cite{Pol}.
 This work arose from a common
project with O. Babelon and M. Talon \cite{ABT}. I also wish to thank the
University of Montreal
and Clarkson University for their hospitality, J. Harnad and F. Nijhoff for
 discussions on this subject. Finally I wish to thank the referee for fruitful
suggestions.

\end{document}